\documentclass[9pt]{article}
\usepackage{spconf,amsmath,graphicx}
\usepackage{multirow}
\usepackage{array}
\usepackage{booktabs}
\usepackage{pifont}
\newcolumntype{P}[1]{>{\centering\arraybackslash}p{#1}}
\usepackage{xcolor}

\title{A study of the robustness of raw waveform based speaker embeddings under mismatched conditions}
\name{Ge Zhu, Frank Cwitkowitz and Zhiyao Duan\thanks{This work is partially funded by Voice Biometric Groups and National Science Foundation grant DGE-1922591.}}
\address{University of Rochester, Rochester, NY, USA \\
         Department of Electrical and Computer Engineering \\
         \small{\tt \( \{ \)gzhu6,fcwitkow\( \} \)@ur.rochester.edu \text{} zhiyao.duan@rochester.edu}}
\begin{document}
%
\maketitle
\begin{abstract}
In this paper, we conduct a cross-dataset study on parametric and non-parametric raw-waveform based speaker embeddings through speaker verification experiments. In general, we observe a more significant performance degradation of these raw-waveform systems compared to spectral based systems. We then propose two strategies to improve the performance of raw-waveform based systems on cross-dataset tests. The first strategy is to change the real-valued filters into analytic filters to ensure shift-invariance. The second strategy is to apply variational dropout to non-parametric filters to prevent them from overfitting irrelevant nuance features.
By combining these strategies, we achieve results comparable to spectral based systems on both the VoxCeleb and VOiCEs datasets. Futhermore, we demonstrate that the learned filters carry little noise compared to existing non-parametric learnable front-ends. 
\end{abstract}
\section{Introduction}
\label{sec:intro}
The design and analysis of hand-crafted features inspired by human auditory perception, such as mel frequency cepstral coefficients (MFCCs), has long been an active area of research in audio processing. In recent years, increasing attention has been directed toward the substitution of such features for data-driven raw-waveform models. Earlier research on sample-level deep neural networks (DNNs) has demonstrated the ability to learn suitable feature embeddings directly from the raw waveform for phone classification~\cite{tuske2014acoustic}, music classification~\cite{kim2018sample} and speaker recognition~\cite{muckenhirn2018towards}. The performance of these systems is comparable to, and in some cases even surpasses, traditional spectral based methods. On feature interpretability, Tuske~\textit{et. al.}~\cite{tuske2014acoustic} showed that DNNs are able to learn bandpass filters purely from the raw waveform without any prior knowledge, and that the first layer can be interpreted as performing a ``quasi time-frequency'' analysis on 
audio.

Inspired by these findings, contemporary raw-waveform models typically comprise a modular structure~\cite{jung2020improved,kong2020panns,han2021time,zhu2021vector}: First, a waveform encoder is used to learn a meaningful representation for audio waveforms and to reduce the dimensionality of feature maps, also referred to as `wavegrams'~\cite{kong2020panns}. 
Then, an additional backbone network further processes the wavegram into embeddings. Under this framework, the trainable front-end filterbanks are the key components of raw-waveform based models. Ideally, the filters should only model task-relevant information, while ignoring other nuisance aspects~\cite{loweimi2020robustness}. However, directly learning from densely sampled audio inputs using DNNs without any prior knowledge can lead to over-fitting~\cite{zeghidour2021leaf}. 

There are two main strategies for effectively training a group of meaningful filters from scratch to achieve comparable results to spectral features. These include parametric filterbanks, and non-parametric filterbanks combined with some initialization or regularization strategy. Both filterbank variations are trained together with a respective network architecture. Learnable parametric filterbanks constrain the filters by optimizing only a few parameters, \textit{e.g.}, center frequency and bandwidth~\cite{zeghidour2021leaf, ravanelli2018speaker}, of pre-defined parametric functions. With such strong constraints, the learned filters generally follow expected shapes and are easier to interpret. 
In contrast, non-parametric learnable filterbanks have little to no regularization. 
In order to mediate this lack of structure, various techniques borrowed from signal processing, such as Gabor initialization~\cite{zeghidour2018learning}, multi-scale analysis~\cite{zhu2021vector}, learnable compression functions~\cite{zeghidour2021leaf} and complex convolution~\cite{peng2021icspk}, are usually applied to the first few layers to avoid overfitting and to speed up convergence.

The performance of raw-waveform based models on cross-domain speech recognition~\cite{loweimi2020robustness,agrawal2020interpretable} and source separation~\cite{pariente2020filterbank} tasks is known to be susceptible to mismatch that on in-domain datasets. In this paper, we compare the efficacy of raw-waveform speaker embeddings to that of traditional mel-spectrum based methods under different acoustic conditions. We propose several strategies to improve the performance of raw-waveform embeddings on cross-domain tasks, including making use of filter analycity and variational dropout to learn sparse filter coefficients. Finally, we visualize and analyze the learned filter responses. The complete code for training and inference will also be made available\footnote{https://github.com/gzhu06/TDspkr-mismatch-study}.

\section{Cross-dataset Studies}
\label{sec:cross}
In this section, we present an empirical study comparing several raw-waveform based speaker embeddings with mel-spectrum based models under both matched and mismatched conditions across several speaker verification tasks.

\subsection{Datasets}
\textbf{VoxCeleb}~\cite{Nagrani17VoxCeleb1,Chung18bVoxCeleb2} is a large-scale dataset containing speech spanning a wide range of speakers under uncontrolled acoustic conditions. We use the VoxCeleb2 development partition for training. We also add 100k augmented noisy utterances by adding reverberation, noise, music, and babble to the original speech files following the Kaldi~\cite{Povey11TheKaldi} recipe\footnote{https://github.com/kaldi-asr/kaldi/tree/master/egs/voxceleb/v2}. We use the full VoxCeleb1 dataset, including Vox1-O, Vox1-E and Vox1-H, to perform matched condition tests. 

\textbf{VOiCEs}~\cite{richey2018voices}, i.e., the Voices Obscured In Complex Environmental Settings corpus, was released with the aim to simulate realistic data under complicated acoustic conditions. It was created by playing Librispeech~\cite{Panayotov2015ls} recordings inside multiple room configurations and re-recording with 12 different microphones placed at various locations. In addition, pre-recorded background noise plus reverberation or echo were played along with the foreground speech. For evaluating the robustness under mismatched conditions, we used the evaluation partition of this corpus, which consists of 3.6 million trial pairs derived from 11,392 utterances.

\subsection{Experimental setup}
\label{ssec:exp1}
We select one parametric waveform encoder, SincNet~\cite{ravanelli2018speaker}, and two non-parametric encoders, multi-scale filters~\cite{zhu2021vector} and TDFbank~\cite{zeghidour2018learning}, to compare against mel-spectrum based system. All of the speaker embedding systems employ 30 filters of length 400 sample (25ms sampling at 16kHz) with a stride size of 5 to extract speech features. Then we feed the output of these three trainable filterbanks to the same backbone network. We model the common backbone with sample-level CNN architectures~\cite{kim2018sample,kong2020panns,zhu2021vector}. Specifically, the waveform embeddings output from the learnable filters are first fed into five down-sampling blocks with a decimation rate of 2. Hence, the sequence length of the feature maps is reduced by a factor of 160 in total, equating to 10 ms of hop size. In the downsampling block, we replace the original dense convolution in~\cite{zhu2021vector} with simple depth-wise separable convolutions, inspired by~\cite{han2021time,luo2019conv}. In this way, the number of parameters is largely reduced. Finally, speaker embeddings are extracted with time delay neural networks (TDNN). 

For the spectral baseline, we use fixed mel-scaled filterbank and the above mentioned backbone network, named `x-conv-vector', for a fair comparison. As a sanity check of the TDNN model's capability, we also train a vanilla MFCC based model, `x-vector (Kaldi)' and a mel-fbank based model `x-vector' in PyTorch for reference.
In order to eliminate the influence of back-end scoring systems on the final verification results, we simply used cosine similarity for scoring. We also compute the equal error rate (EER) to compare different systems.



\subsection{Results and analysis}
\label{ssec:res1}
\begin{figure}[!h]
\centering
\includegraphics[width=2.3in]{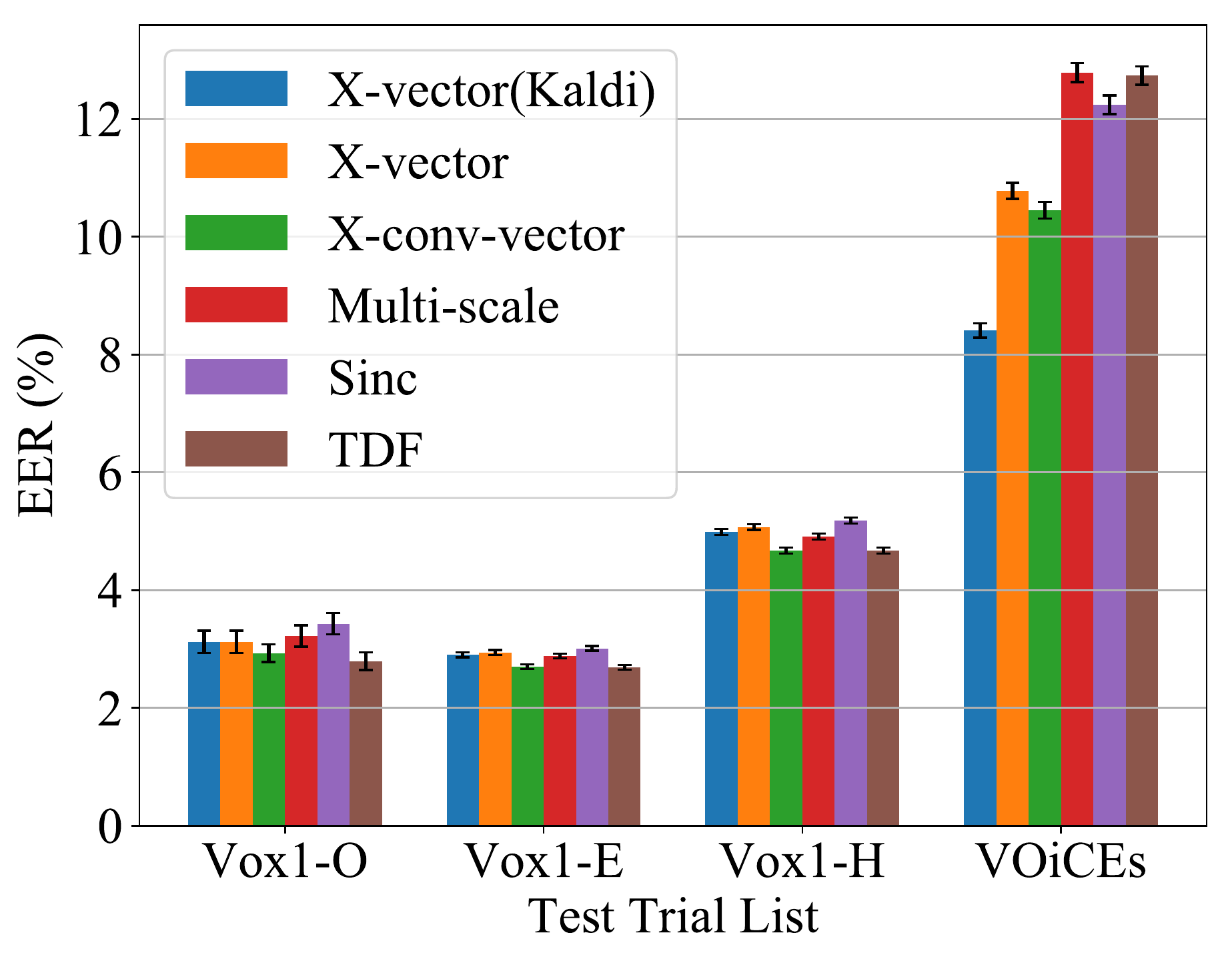}
\caption{EER ($\%$) comparison of mel-spectrum based models and raw-waveform based models on different test sets with cosine similarity scoring. Error bars show a 95$\%$ confidence interval.}
\label{fig:compeer}
\vspace{-2mm}
\end{figure}
In Fig.~\ref{fig:compeer}, we demonstrate EER degradation across datasets for raw-waveform based speaker embeddings. In matched test conditions on VoxCeleb datasets, raw-waveform based speaker embeddings perform on par with the three mel-spectrum based systems. However, in the VOiCEs evaluation dataset, both parametric and non-parametric waveform models lead to degradation compared to spectral based models. It is noted that among all of the six methods, only `x-vector (Kaldi)' performs voice activity detection, making it a less fair baseline. This may also be an important reason for the performance mismatches among mel-spectrum baselines.

\begin{figure}[!t]
\centering
\includegraphics[width=2.9in]{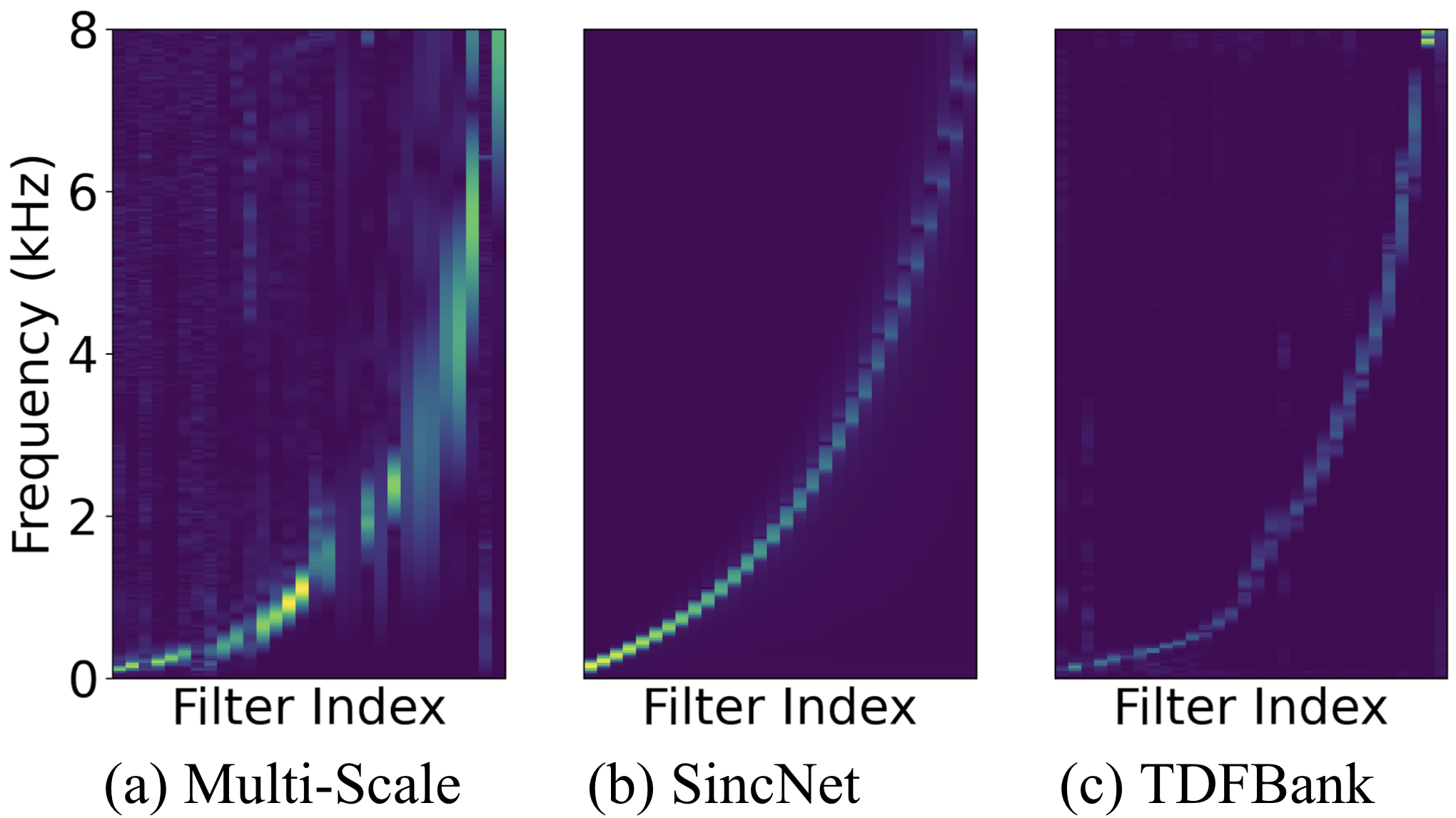}
\caption{Learned filter responses (normalized by the maximum value for better visualization): (a) multi-scale filters, (2) sinc filters (3) the real part of TDFbank.}
\label{fig:compfilter}
\vspace{-4mm}
\end{figure}

We also visualize learned filter responses after training on the noise augmented VoxCeleb dataset, shown in Fig.~\ref{fig:compfilter}. We can see that multi-scale filters and TDFbank are much noisier compared to SincNet, and the frequency resolution in the higher frequencies of multi-scale filters is worse. 

\section{Robust improvement strategies}
\label{sec:method}
In this section, we propose two strategies and discuss their effect on the robustness of raw-waveform speaker embeddings under mismatched conditions. Neither strategy introduces additional parameters or computation. The experimental settings and training details are the same as in section~\ref{ssec:exp1}, except one thing: we also integrated PLDA scoring for the final comparisons on complete speaker verification systmes. We adopt the Gaussian PLDA from Kaldi, which was trained on the augmented VoxCeleb-2 training dataset and evaluated on both the VoxCeleb1 and VOiCEs test datasets. Before training, the extracted speaker embeddings were projected onto a 200-dimensional vector with LDA, followed by whitening and length normalization. 
\vspace{-1mm}
\subsection{Proposed method}
\textbf{Analytic filterbanks.} In the original TDFbank architecture, $N$ real filters and $N$ imaginary filters are initialized into analytic pairs with Gabor wavelets to approximate the mel-filterbanks. Then, a magnitude response is computed using L2 pooling across the output of the real and imaginary pairs. Under the original setting, the weights of the real and imaginary filter components are independently trained without any constraints. As a result, although the initial mel-scale of frequency is mostly preserved after training,the analyticity of the initialization is not preserved. Analytic filters~\cite{flanagan1980parametric} are shift-invariant with respect to time, a desirable property for time-frequency representations. Downsampled convolutions or pooling layers in waveform encoders are not shift-invariant, which compromises their performance on robust classification tasks~\cite{zhang2019making}. A natural way to constrain the analycity of learned complex filterbanks is to learn only the real component of a filter, and to and infer the imaginary component directly using the Hilbert transform~\cite{pariente2020filterbank, cwitkowitz2021learning}. 
In this way, the magnitude of the filter response is shift-invariant and the number of filter parameters is essentially halved. Therefore, in this work, we apply the Hilbert transform to obtain the corresponding imaginary filters of real filters. We do this for both the non-parametric and parametric sinc filters. 

\textbf{Sparse variational dropout.} Observing the noisy filter responses in Fig.~\ref{fig:compfilter}, we believe that the non-parametric filters tend to overfit the noisy training data, learning nuisance aspects of the recordings. One way to ease this problem is to regularize the network 
by dropping irrelevant weights with sparse variational dropout (VD)~\cite{molchanov2017variational}. VD was originally proposed as a model compression technique to sparsify DNN weights. In this work, we follow our previous work~\cite{cwitkowitz2021learning} to sparsify filters by applying VD in the first layer of the raw-waveform models. 

Dropout can be seen as injecting fixed Bernoulli noise or Gaussian noise into weights during training. 
Instead of setting a fixed variance as in Gaussian dropout (GD), VD injects an individual multiplicative Gaussian noise $\xi_{ij}\sim N(1,\alpha_{ij})$ to every weight, with the variance $\alpha_{ij}$ consisting of model parameters learned with an approximated KL-divergence measure. By learning an individual variance for every weight, VD is able to induce sparsity across learned weights when $\alpha_{ij}\to\infty$ (equivalent to $p=1$ in Bernoulli dropout). In such cases, the weights can be ignored or removed from neural networks during inference time.  

\begin{table*}[th]
\caption{EER (\%) comparison on different test sets. All models are trained on the noise augmented VoxCeleb2 training set and scored with PLDA backend. A statistical significance test is performed using a bootstrap procedure~\cite{haasnoot18eerci}: an absolute value of 0.05 of EER difference for Vox1-E and Vox1-H is outside the 95$\%$ confidence interval for all methods, while for Vox1-O and VOiCEs the EER difference has to be larger than 0.15 and 0.13 respectively.}  
\setlength{\tabcolsep}{6pt}
\renewcommand{\arraystretch}{0.86}
\centering
\begin{tabular}{P{80pt}P{43pt}P{25pt}P{40pt}P{25pt}P{40pt}P{25pt}P{40pt}P{25pt}P{40pt}}
\toprule
\multirow{2}{*}{\textbf{System}} &\multirow{2}{*}{\textbf{Feature}} &
\multicolumn{2}{c}{\textbf{VoxCeleb-O}} &
\multicolumn{2}{c}{\textbf{VoxCeleb-E}} &
\multicolumn{2}{c}{\textbf{VoxCeleb-H}} &
\multicolumn{2}{c}{\textbf{VOiCEs}}\\ 
\cline{3-10}
&&EER&min-DCF&EER&min-DCF&EER&min-DCF&EER&min-DCF\\
\hline
x-vector (Kaldi) &MFCC&2.26 & 0.256 &2.37&0.279 &4.14&0.408&\textbf{6.79}&\textbf{0.553}\\ 
x-vector &Mel-fbank&2.37 &0.264  &2.42&0.280&4.18 &0.406&8.14&0.658\\ 
x-conv-vector&Mel-fbank&\textbf{2.04} &\textbf{0.241}  &\textbf{2.17}&\textbf{0.252}&\textbf{3.79} &\textbf{0.379}&7.10&0.581\\
\hline
Multi-scale&\multirow{7}{*}{Waveform}&2.28&0.273   &2.38&0.285& 4.17&0.408&8.54&0.705 \\ 
Sinc &&2.37 &0.287 &2.32  &0.278&4.02&0.400 &8.55&0.682\\ 
\textbf{Sinc+$\mathcal{H}$} &&2.15 &0.270 &2.28  &0.271&3.91&0.396 &8.90&0.669\\ 
TDF &&\textbf{1.98} &\textbf{0.230} &\textbf{2.19}  &\textbf{0.249}&\textbf{3.85}&\textbf{0.383}  &8.38&0.663\\ 
\textbf{TDF+$\mathcal{H}$}&&2.01&0.261&2.27&0.263&3.98&0.396&7.46&\textbf{0.621}  \\ 
\textbf{TDF+VD}&&1.98 &0.235   &2.30 &0.264 &4.05 &0.385 & 7.68& 0.626 \\ 
\textbf{TDF+$\mathcal{H}$+VD}&&1.99&0.266  &2.26&0.253&3.93&0.385&\textbf{7.40}&0.633  \\ 
\bottomrule
\end{tabular}
\label{tab:results}
\vspace{-2mm}
\end{table*}

\vspace{-1.5mm}
\subsection{Results}
\textbf{Comparison.} In this experiment, we evaluate the proposed strategies in the same experimental setup as in Section~\ref{sec:cross}. 
We can see that the `Multi-scale', `Sinc' and `TDF' baselines in Table~\ref{tab:results} show more degradation on VOiCEs test set compared to spectral baselines, which is consistent with the conclusion in Sec.~\ref{sec:cross}. By comparing `TDF' and `Sinc' with their corresponding analytic versions, we find that `Sinc+$\mathcal{H}$' only achieves a marginal improvement over the `Sinc' baseline on VoxCeleb but a slight degradation on VOiCEs, whereas `TDF+$\mathcal{H}$' significantly outperforms the `TDF' baseline on VOiCEs and yields comparable results on VoxCeleb. This shows that the analyticity constraint helps non-parametric filters learn robust representations, but it is not the case for parametric filters. This may be because the benefit of filter analyticity is mainly on learning transient components, which cannot be well modeled in the sinc filters anyway.
Comparing `TDF' and `TDF+VD', we can also observe a significant improvement on VOiCEs without compromising the performance on VoxCeleb with the help of VD.
Among all of the raw-waveform based systems, `TDF+$\mathcal{H}$+VD' achieves the best results on the out-of-domain test set, with both VD and analytic filters helping to boost the performance. Compared with the three spectral based models, it achieves comparable results to `x-conv-vector' with a similar model size and training strategy. 
Note that x-vector (Kaldi) achieves similar performance to that reported in ~\cite{lin2020learning} on the VOiCEs dataset; this further validates our TDNN backbone implementation. 

\begin{table}[h]
\caption{EER (\%) comparison on different test sets. All models are trained on the augmented VoxCeleb2 training set and scored with cosine similarity.} 
\setlength{\tabcolsep}{3pt}
\renewcommand{\arraystretch}{0.84}
\centering
\begin{tabular}{c|cccc}
\toprule
\multicolumn{1}{c|}{System}   & \multicolumn{1}{c}{Vox1-O} & \multicolumn{1}{c}{Vox1-E} & \multicolumn{1}{c}{Vox1-H} & \multicolumn{1}{c}{Voices} \\ \hline
x-vector (Kaldi)&	3.12&	2.9	&4.99&	8.41 \\ 
x-vector	&3.12&	2.94&	5.07&	10.78 \\ 
x-conv-vector	&2.93&	2.7	&\textbf{4.67}&	10.45 \\\hline
TDF  & 2.79& \textbf{2.69}& 4.67 & 12.74 \\  
TDF+VD &3.01&2.79&4.81&11.10\\ 
TDF+$\mathcal{H}$ &\textbf{2.72}&2.81&4.86&10.72\\ 
TDF+$\mathcal{H}$+BD & 3.06& 2.77& 4.83 & 11.69\\ 
TDF+$\mathcal{H}$+GD&2.98  &2.73 & 4.83  & 11.29 \\ 
TDF+$\mathcal{H}$+VD & \textbf{2.72} & 2.72   & 4.72& \textbf{10.32}\\ 
 \bottomrule
\end{tabular}
\label{tab:abstudy}
\vspace{-3mm}
\end{table}

\textbf{Ablation study.} In order to better demonstrate the effectiveness of each component without the influence of the scoring backend, we conducted several ablation studies using cosine similarity, as shown in Table~\ref{tab:abstudy}. The improvement of applying analytic filters is consistent with the PLDA backend results in Table~\ref{tab:results}. Different from results in Table~\ref{tab:results}, we find that `TDF+$\mathcal{H}$+VD' outperforms `TDF+$\mathcal{H}$' when cosine similarity is used. Similarly, `TDF+$\mathcal{H}$+VD' outperforms `x-conv-vector' slightly on VOiCEs. These differences suggest that by dropping filter weights through VD, the final learned speaker embeddings tend to become less Gaussian, hence yield worse results with the PLDA backend. 
We also experimented with different dropout techniques shown in the last four rows in Table~\ref{tab:abstudy}, we can observe that BD and GD are not helpful in improving robustness compared to `TDF+$\mathcal{H}$' baseline, while VD achieves better verification results in all of in-domain and out-of-domain tasks.

\begin{figure}[!h]
\centering
\includegraphics[width=3.3in]{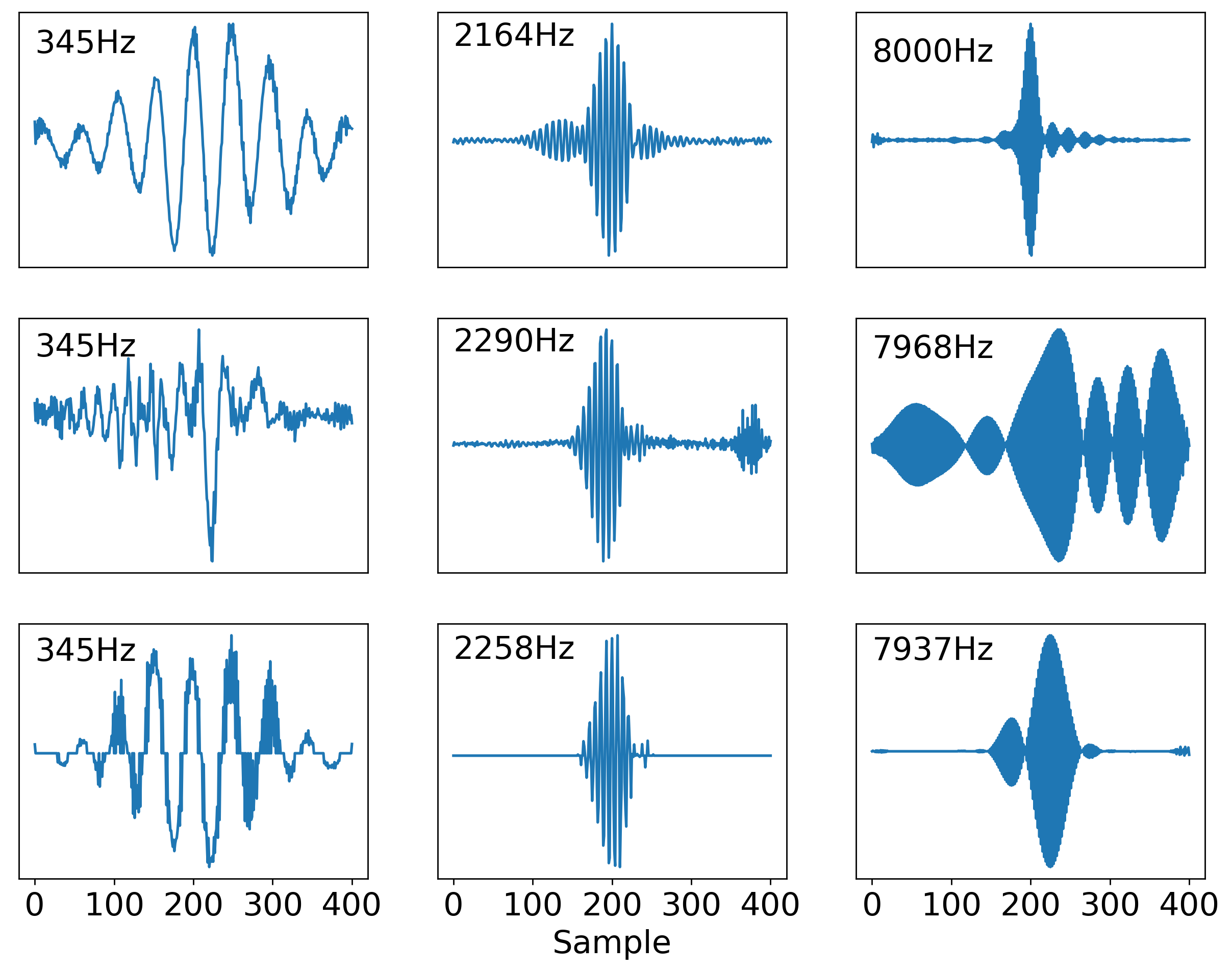}
\caption{Examples of learned filters with their maximum response frequency labeled. Top row: `TDF+$\mathcal{H}$' filters trained on clean VoxCeleb. Middle row: `TDF+$\mathcal{H}$' filters trained on noise augmented VoxCeleb. Bottom row: `TDF+$\mathcal{H}$+VD' filters trained on noise augmented VoxCeleb.}
\label{fig:compfilterstime}
\vspace{-3mm}
\end{figure}

\textbf{Filter visualization.}
In Fig.~\ref{fig:compfilterstime}, we visualize several learned non-parametric filters at different frequency bands under different training settings for TDF based methods. 
When trained on the noisy dataset, the learned filters are less regular and much noisier than filters trained on the clean dataset. With the help of VD, the learned filter at 345Hz is similar to the one trained without noise, and only the center weights of the filters at 2258Hz and 7937Hz are retained. The `jitters' picked up from the noise are not present in the filters. Although there is no significant improvement on EER over the baseline with VD, this verifies that during training, raw waveform models tend to capture nuisance information from noisy data, and proves that dropping out the corresponding weights does not affect the final performance.

\section{Conclusion}
\label{sec:conclusion}
In this paper, we performed a systematic empirical study of multiple parametric and non-parametric raw-waveform based speaker embeddings. 
In comparison to several mel-spectrum baselines, these raw-waveform based methods yield similar results on in-domain tests, but show a more significant degradation on cross-domain tests.
In order to bridge this performance gap, we proposed to apply filter analyticity to promote shift-invariance of the learned filters and variational dropout on non-parametric filters to discard task irrelevant information during training. Finally, we observed a significant improvement for non-parametric raw-waveform based embeddings with respect to cosine similarity and PLDA backends, achieving similar performance to the mel-spectrum baselines.



\vfill\pagebreak

\small
\bibliographystyle{IEEEbib}
\bibliography{refs}

\end{document}